# Dual camera snapshot hyperspectral imaging system via physics-informed learning

### HUI XIE, ZHUANG ZHAO,* JING HAN,* YI ZHANG, LIANFA BAI AND JUN LU

*School of Electronic Engineering and Optoelectronic Technology, Nanjing University of Science and Technology, Nanjing 210094, China*
*\*Corresponding author: Zhaozhuang@njust.edu.cn, eohj@njust.edu.cn*

**Abstract:** We consider using the system's optical imaging process with convolutional neural networks (CNNs) to solve the snapshot hyperspectral imaging reconstruction problem, which uses a dual-camera system to capture the three-dimensional hyperspectral images (HSIs) in a compressed way. Various methods using CNNs have been developed in recent years to reconstruct HSIs, but most of the supervised deep learning methods aimed to fit a brute-force mapping relationship between the captured compressed image and standard HSIs. Thus, the learned mapping would be invalid when the observation data deviate from the training data. Especially, we usually don't have ground truth in real-life scenarios. In this paper, we present a self-supervised dual-camera equipment with an untrained physics-informed CNNs framework. Extensive simulation and experimental results show that our method without training can be adapted to a wide imaging environment with good performance. Furthermore, compared with the training-based methods, our system can be constantly fine-tuned and self-improved in real-life scenarios.

## 1. Introduction

Hyperspectral imaging techniques can provide tens to hundreds of discrete bands of electromagnetic reflection-based real scenes [1]. The spectral details in hyperspectral imaging showed the confirmed information about the illumination and materials, which is beneficial to several domains such as material classification, remote sensing, and pathological examination [2-6]. Therefore, hyperspectral imaging technology has attracted considerable attention from academia and industry over the past few decades [7-11].

To overcome the limitation of measurement acquisition time in traditional hyperspectral systems, several new imaging methods have been explored during computational imaging [12-15] development. In these works, coded aperture snapshot spectral imaging (CASSI) [16], including dual-disperser architecture (DD-CASSI) [17] and single disperser design (SD-CASSI) [18], has attracted extensive attention from researchers. However, because the unconstrained optimization algorithms problem that is required to be solved by CASSI is undetermined and the coded method is single, it is difficult to achieve the high-quality reconstruction requirements. To improve the quality of CASSI imaging, Kitti et al. proposed a multi-frame CASSI (MS-CASSI) [19], which uses multiple different coded apertures to capture the same scene. However, multi-frame CASSI lacked its most prominent feature snapshot function; therefore, it is not suitable for dynamic scenarios. Arce et al. proposed a colored coded aperture compressive spectral imaging (CC-CASSI) [20] system to replace the traditional blocking–unblocking coded apertures and extended the compressive capabilities of CASSI. Lin et al. proposed spatial–spectral-encoded compressive hyperspectral imaging (SS-CASSI) [21], and a flexible capture mode using a dual-coded compressive hyperspectral imaging system (DCSI) was proposed by Wang et al. [22]. The optical structures of these systems are shown in Fig. 1.

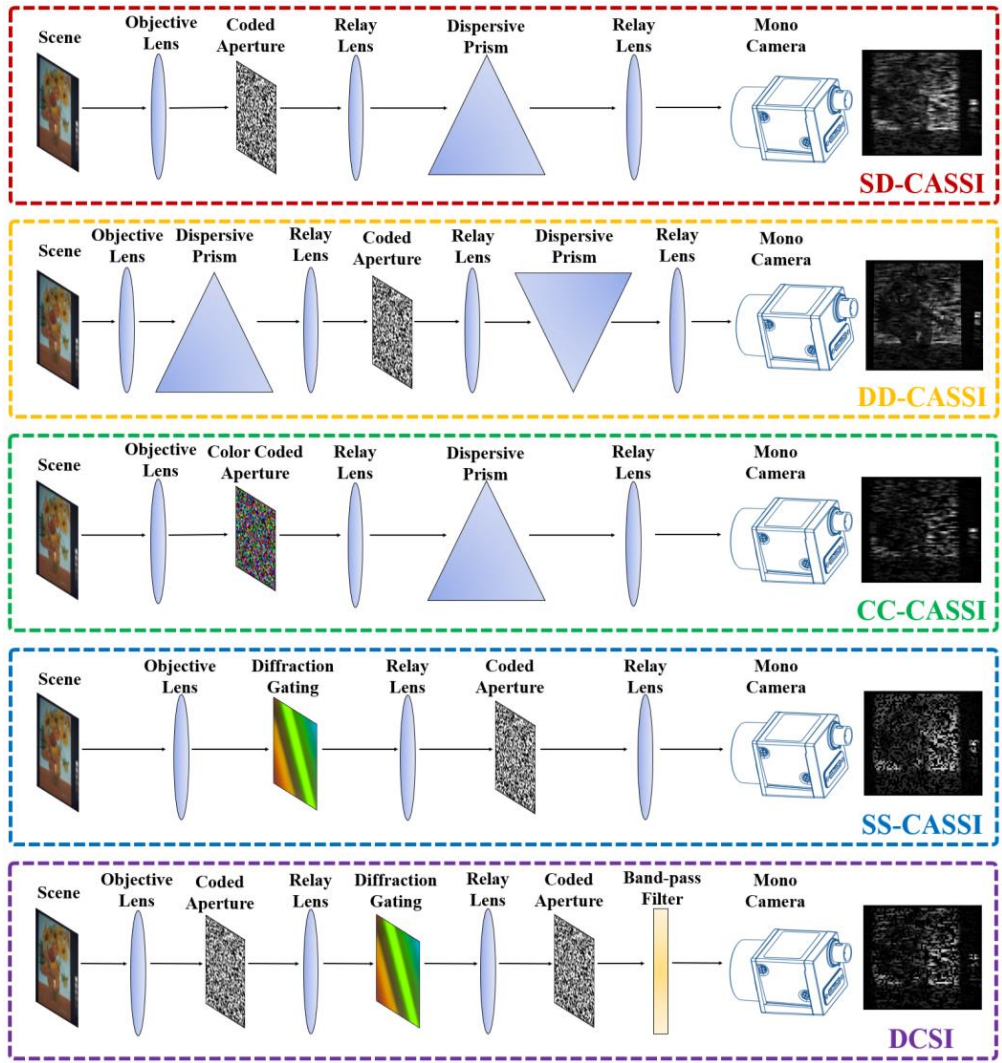

Fig. 1 Illustration of representative imaging systems.

In recent years, hyperspectral imaging using supervised deep-learning methods [8-10] [23-27] has attracted increasing attention. Two advanced convolutional neural networks for the hyperspectral reconstruction mission, collectively known as HSCNN+, were proposed by Xiong et al. [28]. Miao et al. established λ-net for compressive-spectral imaging reconstruction, which can perform the reconstruction task within sub-seconds [7]. Fu et al. reported an end-to-end deep learning method for reconstructing hyperspectral images from a raw mosaic image [8].

However, amount of information is lost in the spectral integration process when sensors capture light. Most of the well-known supervised CNNs methods attempt to fit the mapping relationship between the compressive images and standard HSIs. To solve the problem of learning an accurate prior in HSIs, Wang et al. upgraded CASSI to a dual-camera design and presented a CNNs method from external and internal learning to guarantee the generalization ability [22]. But hyperspectral datasets are inadequate for representing the diversity of real-world scenarios, it might limit the generalizability of existing CNNs methods. There for, Meng et al. developed an untrained neural network for HSIs reconstruction by integrating deep image priors into the plug-and-play regime [24].

In this paper, we used the system's optical imaging process with untrained CNNs to solve the snapshot hyperspectral imaging reconstruction problem. To this end, we built a physics-informed self-supervised CNNs framework with the spectral quantum efficiency curves of the color camera, and the optical imaging process of SD-CASSI [18]. And we built a dual-camera system to simultaneously capture both the color image of the scene and the encoded compressed image. Our system utilized the CNNs to reconstruct three-dimensional HSIs from the two-dimensional compressive images directly and doesn't need any training data.

## 2. Methods

The principle of snapshot compressive imaging (SCI) system is to encode the high-dimensional data onto a two-dimensional measurement. CASSI system is one of the earliest SCI systems to capture the HSIs cube in a snapshot way. As shown in Fig. 2. We placed an optical beam splitter coating in front of the SD-CASSI system and used a color camera to collect scene data for dual-camera equipment construction.

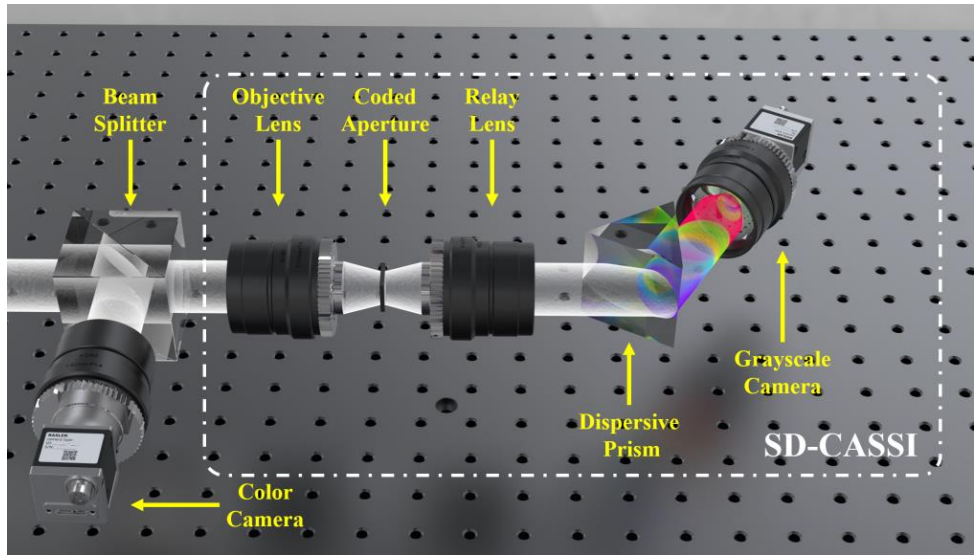

Fig. 2. Schematic illustration of our dual-camera equipment.

In the spatial modulation of the CASSI branch, the scene information is first projected onto the coded aperture. Thereafter, the spatially modulated information is dispersed through the dispersion prism and captured by the camera. Mathematically, the information of three-dimensional HSIs can be expressed as $h(x, y, \lambda)$, where $(x, y)$ are the space coordinates ($1 \leq x \leq X, 1 \leq y \leq Y$) and $\lambda$ indexes the spectral coordinate ($1 \leq \lambda \leq N$).

For the color camera sensor, the raw image received by the color camera can be expressed as:

$$f_{\text{color}}(x, y) = \sum_{\lambda=1}^{N} h(x, y, \lambda) \, K(x, y, \lambda) L(\lambda) \tag{1}$$

where $K(x, y, \lambda)$ represent the spectral quantum efficiency of the color camera color filter array, $L(\lambda)$ represents the illumination spectrum, and N represents the spectrum channel number.

For the CASSI branch, the spectral density modulated by the coded aperture and after the dispersive prism, the compressive image finally acquired by the grayscale camera can be related as:

$$f_{grayscale}(x, y) = \sum_{\lambda=1}^{N} h(x - \lambda, y, \lambda) T(x - \lambda, y) \qquad (2)$$

where $T(x, y)$ represents the coded aperture. The coded aperture was first reshaped into a column vector, then it was uniformly repeated as the number of spectral bands evenly in the horizontal direction, and zeros were filled in the vertical direction, as shown in Fig. 3.

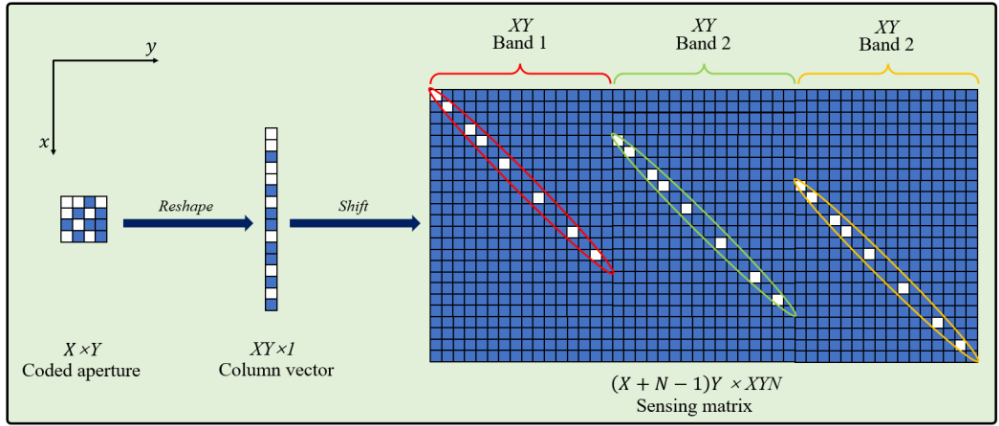

Fig. 3. Relationship between the coded aperture and the sensing matrix of the camera.

Hence, Eq. 2 can be written as

$$F = \Phi H \qquad (3)$$

where $F \in \mathbb{R}^{(X+N-1)Y}$ and $H \in \mathbb{R}^{XY}$ are the vectorized representations of the compressive image $f_{grayscale}(x, y)$ and the HSIs $h(x, y, \lambda)$, and $\Phi$ is the physical process of the CASSI system, which is transformed by $T(x, y)$. For the CASSI system, if we use a three-dimensional HSI patch of size $X \times Y \times N$ through the system, the size of the finally collected two-dimensional compressive image is $(X + N - 1) \times Y$, as shown in Fig. 4.

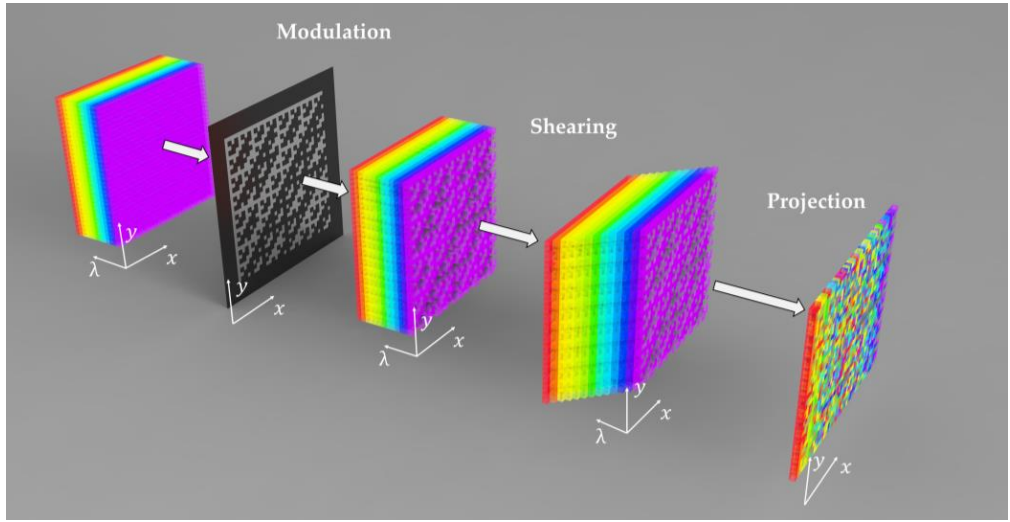

Fig. 4. Sensing matrices of multispectral acquisition.

In this study, we aimed to reconstruct HSIs $h(x, y, \lambda)$ from two compressive images $f_{color}(x, y)$ and $f_{grayscale}(x, y)$.

Exploring the best solution, we present a self-supervised learning system with an untrained convolutional neural network (CNN) and a dual-camera equipment, as shown in Fig. 5.

(i). First, we design a spectral image reconstruction physics-informed CNN framework based on a parallel-multiscale network [29] to adequately use the spectral correlation between adjacent pixels and that between adjacent bands. It can be understood as a dimensional ascension from a two-dimensional compressive image captured by a grayscale camera.

(ii). Then, we used the reconstructed HSIs to create two self-supervised learning branches. One branch projected the physical process of the CASSI to obtain a compressed image to learn the spectral information. Another branch, along with the quantum efficiency curve of the color camera, emulated the imaging process of the color camera; Thus, we obtain a color image for the network to learn the spatial information.

(iii). The dual-camera equipment with the two self-supervised learning branches formed a closed-loop online self-learning system.

To obtain the optimal parameters of the system, we used the mean square error between the images $p_{color_i}(x, y)$, $p_{grayscale_i}(x, y)$ —projected by self-supervised models—and images $f_{color_i}(x, y)$, $f_{grayscale_i}(x, y)$.—captured by the color and grayscale cameras. The calculations are:

$$\text{Loss}_{color} = \frac{1}{n}\sum_{i=1}^{n} \left| p_{color_i}(x, y) - f_{color_i}(x, y) \right|^2 \quad (4)$$

$$\text{Loss}_{grayscale} = \frac{1}{n}\sum_{i=1}^{n} \left| p_{grayscale_i}(x, y) - f_{grayscale_i}(x, y) \right|^2 \quad (5)$$

where n represents the total number of self-supervised learning data.

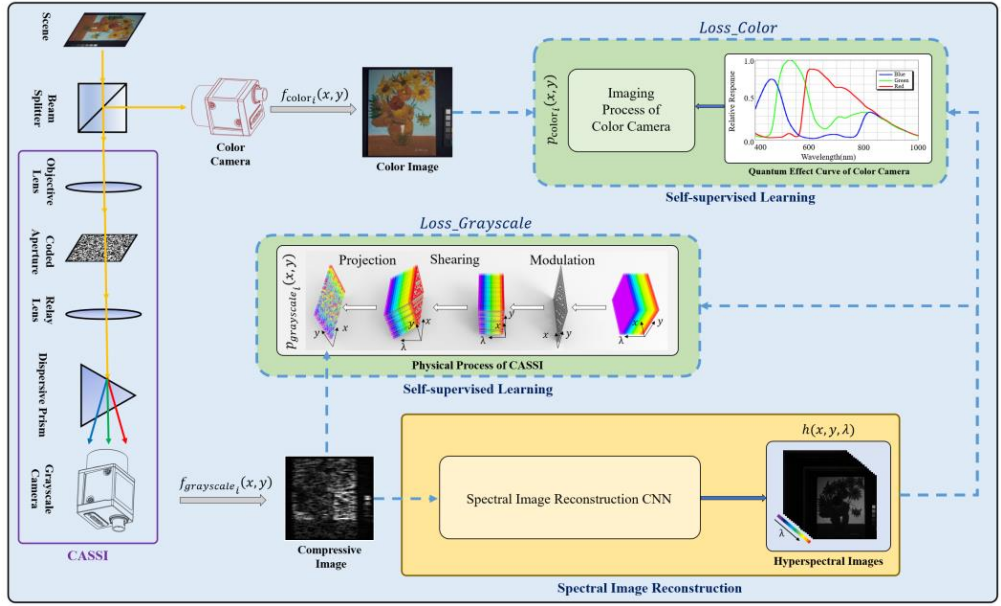

Fig. 5. Schematic illustration of our closed-loop online self-supervised learning system.

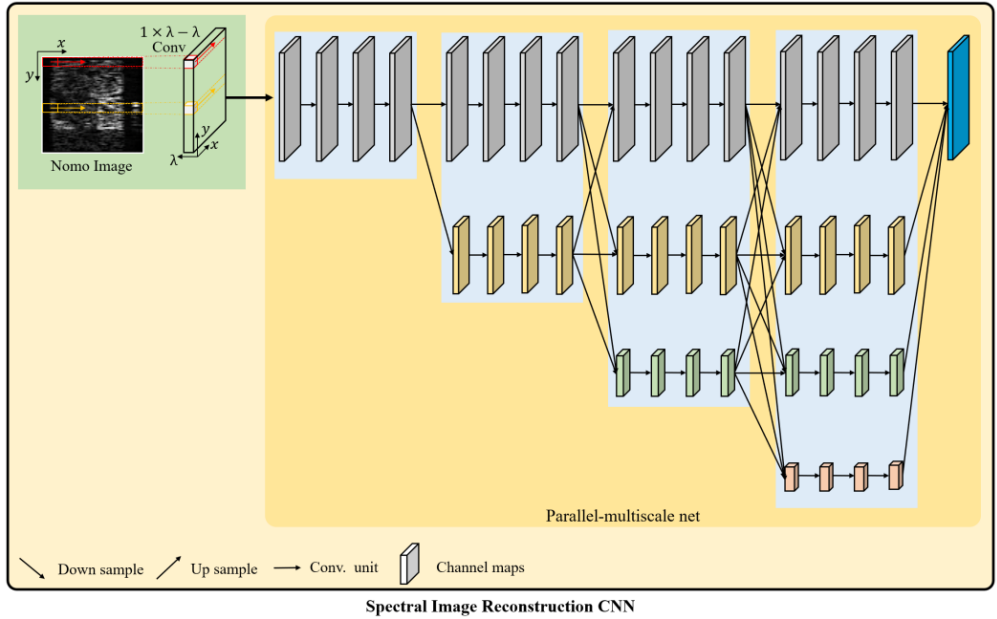

Fig. 6. Spectral image reconstruction CNN. It consists of two stages. The first stage uses λ convolution filters of size 1×λ (λ is the number of HSIs channels) for dimensional ascending. The second stage is the parallel-multiscale network for HSIs completion.

Compared with other supervised learning-based methods [7] [24], we used the self-supervised framework to build the physical imaging process instead of to fit the mapping relationship between the compressive images and standard HSIs. So, we don't require a significant amount of standard data for pre-training. Compared with Meng et al.'s self-supervised network [24], we both used untrained neural networks to solve the reconstruction problem of snapshot compressive imaging. But different from Meng et al.'s alternating optimization algorithm with joint network learning and reconstruction. We considered the constraint of RGB and CASSI measurements with two self-supervised branches. And after the optimization, the learned model has good reconstruction ability for the same type scenarios. So, the learned model can provide a real-time reconstruction capacity, just like the supervised deep learning networks after training.

## 3. Experiments and results

### 3.1 Simulation setup

We used three public hyperspectral datasets, CAVE [30], ICVL [31], and Harvard [32], to verify the self-supervised system. The CAVE dataset consists of 31 indoor scenes illuminated with the CIE standard illuminant D65. The ICVL dataset consists of 201 outdoor scenes. The Harvard dataset includes 50 images of indoor or outdoor images under daylight and 27 indoor images with artificial or mixed illumination. These three datasets contain reflectivity data at full spectral resolution from 400 nm to 700 nm at a step size of 10 nm (total 31 bands). To formulate the datasets, we cropped the images into 256×256 patches.

In our experiments, we set $\lambda=31$ for the network, and the batch size was set to 16. The initial value of the learning rate was set to 0.001 and decreased by a factor of 5 for every 300 epochs; the total number of epochs was 1200. All the tests implemented on Intel Core i7-10700 CPU, 32G RAM and an Nvidia RTX3090 GPU.

### 3.2 Results on standard data

To verify the capability of the system in this study, we compared it with several state-of-the-art methods, including iterative optimization methods and supervised deep learning methods. The performance of these methods was evaluated using the peak signal-to-noise ratio (PSNR), structural similarity index measure (SSIM), and spectral angle mapping (SAM). For each two-dimensional spatial image, we used PSNR and SSIM to calculate and show the spatial quality between the reconstructed HSIs and reference pictures. The larger the PSNR and SSIM values, the better is the performance. For three-dimensional spectral information, SAM regarded the spectrum of each pixel in the image as a high-dimensional vector, which measures the spectral similarity by calculating the angle between the reconstructed HSIs and the standard datasets. The smaller the SAM value, the better is the reconstruction effect.

Our method is compared with five state-of-the-art HSI reconstruction methods on a standard dataset, including two iterative optimization-based methods such as the two-step iterative shrinkage/thresholding (TwIST) [33] and total variation based methods (GAP-TV) [34], two supervised learning based methods such as the $\lambda$-Net [7] and HSCNN+ methods [28]，and one self-supervised network method proposed by Meng et al. [24].

Because we used untrained neural networks to reconstruct HSIs, ground truth (GT) is not required in the fitting process of our network. Although we do not need to distinguish between the training and test datasets, it is necessary to discuss the reconstruction capabilities of our system. Herein, we divided the verification of our methods into two strategies.

**(i)  Compared with iterative optimization algorithm methods**

Benefitting from our physics-informed self-supervised CNN framework, mapping the relationship between the compressed image and the HSIs is not required. We turn it to learn about the optical imaging process of the color camera and CASSI system. Networks with physical constraints are more direct and efficient than iterative processes (The imaging index is higher and the speed is faster). So, in this strategy, we compared with the iterative optimization-based methods (TwIST and GAP-TV) and Meng et al.'s methods. The evaluation results as show in Table 1. Besides, the running time our proposed approach is 5 minutes per graph (256×256 resolution), which much faster than Meng et. al.'s alternating optimization algorithm (about 30 minutes).

**Table 1. Performance Comparison of our method with iterative optimization methods.**

|  | CAVE | | | ICVL | | | Harvard | | |
| --- | --- | --- | --- | --- | --- | --- | --- | --- | --- |
| Method | PSNR | SSIM | SAM | PSNR | SSIM | SAM | PSNR | SSIM | SAM |
| TwIST | 24.983 | 0.814 | 9.186 | 26.435 | 0.833 | 3.276 | 23.018 | 0.814 | 6.973 |
| GAP-TV | 26.125 | 0.839 | **8.713** | 27.462 | 0.854 | 3.923 | 25.849 | 0.842 | 6.586 |
| Meng et. al. | 32.417 | 0.898 | 10.413 | 31.739 | 0.903 | 3.358 | 30.389 | 0.881 | 5.967 |
| **Ours** | **43.877** | **0.961** | 9.052 | **41.265** | **0.977** | **2.041** | **38.712** | **0.965** | **4.364** |

**(ii)  Compared with the supervised deep learning methods**

Different from the iterative optimization method, the learned model of our method also has the reconstruction ability. It is similar to supervised deep learning methods. So, in this strategy, similar to the training and testing strategies of most supervised deep learning methods. We divide the datasets into training sets and test sets. The CAVE dataset, we selected 20 images randomly for training and the 10 images for testing. The ICVL dataset, thereby containing a lot of similar data. We selected 50 images randomly for training and the others 150 images for testing. For the Harvard dataset, we selected 20 outdoor images and 10 indoor images randomly for training and the rest for testing. The evaluation results as show in Table 2. It can be seen that our approach is nothing less than supervised deep learning methods. It implies the capacity of real-time reconstruction which our system can provide.

**Table 2. Same as the general deep learning method test.**

| CAVE | ICVL | Harvard |
| --- | --- | --- |

| Method | PSNR | SSIM | SAM | PSNR | SSIM | SAM | PSNR | SSIM | SAM |
|---|---|---|---|---|---|---|---|---|---|
| HSCNN+ | 33.097 | 0.905 | 11.823 | 31.253 | **0.954** | 2.493 | 29.248 | 0.903 | 6.185 |
| λ-net | 31.128 | 0.892 | 17.549 | 29.853 | 0.912 | 3.632 | 28.279 | **0.946** | 13.249 |
| **Ours** | **39.151** | **0.943** | **11.097** | **36.674** | 0.950 | **2.291** | **34.334** | 0.933 | **5.34** |

Since our method adopts physical constraints, GT is not required. Our method can be used for online learning of any scene and has better adaptability to the scenes. To discuss the generalization, we used cross-dataset validation across three datasets to verify our framework and the comparison supervised deep learning methods. The results are presented in Table 2–4. It can be obviously seen that generalization is terrible on external datasets. The trained model would fail when there is a large difference between the actual and training data. There have been many deep learning methods to ensure the generalization ability of the system, such as the method of [9]. But we usually don't have ground truth in real-life scenarios, making it impossible for most supervised deep learning methods to fine-tune the scenes. In this case, because we used physics-informed untrained network to solve hyperspectral reconstruction problems, our method can fine-tune the scenes. The fine-tune results are also shown in Table 4–5. This implies the efficacy of our dual physical constraints (RGB and CASSI). With the online self-learning way, our system can achieve the capability of fine-tuning itself in real-life scenarios, and then achieve a stronger adaption ability.

**Table 3. Generalization Comparison of our method with other deep learning methods training on the CAVE dataset**

| Method | ICVL | | | Harvard | | |
|---|---|---|---|---|---|---|
| (train on CAVE) | PSNR | SSIM | SAM | PSNR | SSIM | SAM |
| HSCNN+ | 28.598 | 0.836 | 6.776 | 22.295 | 0.520 | 43.822 |
| λ-net | 27.618 | 0.808 | 6.918 | 21.186 | 0.503 | 48.636 |
| **Ours (no fine-tune)** | 29.471 | 0.879 | 6.387 | 23.205 | 0.557 | 37.088 |
| **Ours (fine-tune)** | **42.398** | **0.983** | **2.135** | **38.819** | **0.969** | **4.107** |

**Table 4. Generalization Comparison of our method with other deep learning methods training on the ICVL dataset**

| Method | CAVE | | | Harvard | | |
|---|---|---|---|---|---|---|
| (train on ICVL) | PSNR | SSIM | SAM | PSNR | SSIM | SAM |
| HSCNN+ | 26.100 | 0.619 | 28.717 | 23.657 | 0.574 | 32.088 |
| λ-net | 19.983 | 0.549 | 30.198 | 21.796 | 0.512 | 36.048 |
| **Ours (no fine-tune)** | 27.218 | 0.690 | 29.882 | 24.446 | 0.590 | 30.664 |
| **Ours (fine-tune)** | **43.914** | **0.969** | **8.131** | **38.918** | **0.971** | **4.112** |

**Table 5. Generalization Comparison with other deep learning methods training on Harvard dataset**

| Method | CAVE | | | ICVL | | |
|---|---|---|---|---|---|---|
| (train on Harvard) | PSNR | SSIM | SAM | PSNR | SSIM | SAM |
| HSCNN+ | 20.231 | 0.613 | 55.467 | 21.451 | 0.629 | 31.769 |
| λ-net | 18.374 | 0.549 | 56.894 | 20.489 | 0.594 | 33.654 |
| **Ours (no fine-tune)** | 22.069 | 0.618 | 44.341 | 23.076 | 0.683 | 26.200 |
| **Ours (fine-tune)** | **44.009** | **0.973** | **8.043** | **41.697** | **0.984** | **2.011** |

To provide the visual comparison, we chose three different scenarios among the ICVL, Harvard, and CAVE datasets, as shown in Fig. 7. The average absolute errors of the spectra between the standard images and reconstructed results are shown in the error maps. We also selected four spatial locations from four scenes to exhibit exemplar spectrum reconstruction, as shown in Fig. 8. To visually compare all methods, we selected three representative scenes from these three datasets, as shown in Fig. 9. It can be seen that our system accomplishes visually pleasant results with high-fidelity reconstruction.

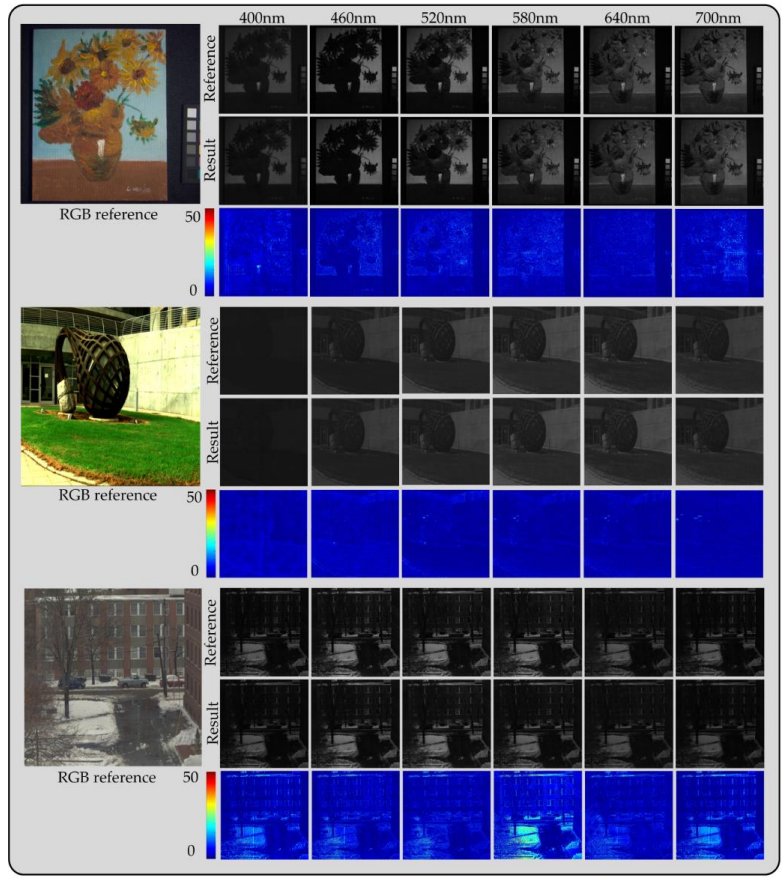

Fig. 7. Image quality comparison of three representative images from the CAVE, ICVL, and Harvard datasets. Six selected channels error maps by using our system.

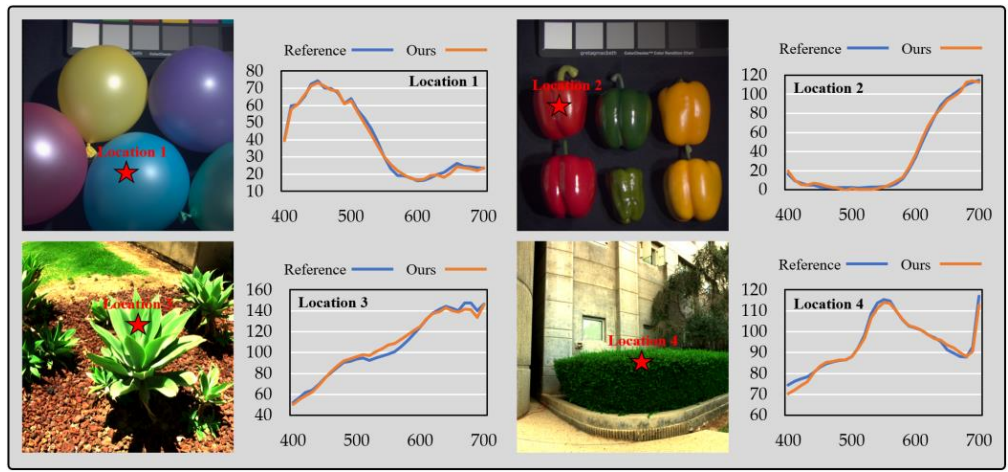

Fig. 8. Reconstructed spectrum of four selected spatial locations from four scenes on the Optimization algorithm strategy. The x-axis represents wavelength (nm) and the y-axis represents spectral intensity.

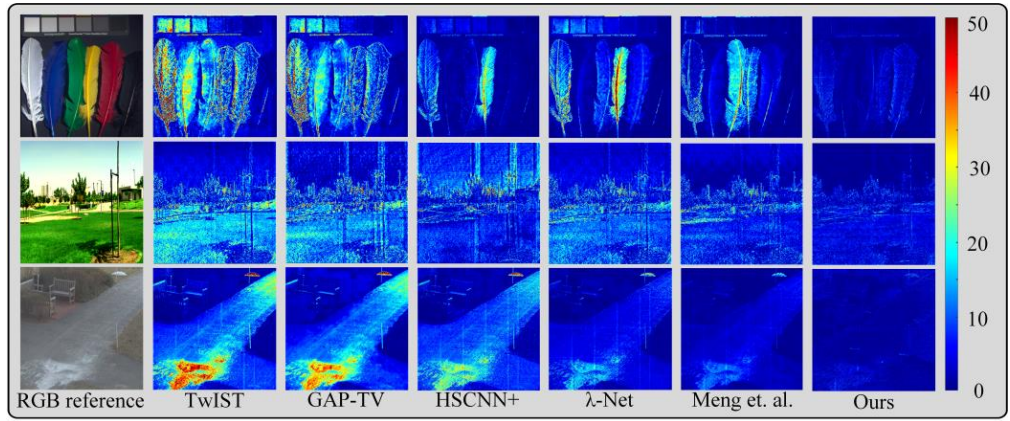

Fig. 9. Comparison of image quality of three representative scenes from three datasets. From left to right are the reconstructed results' error maps of TwIST/GAP-TV/HSCNN+/λ-net/ Meng et. al./Our method (optimization algorithm), and the corresponding color reference are provided on far left.

### 3.3 Experimental results

We also conducted such simulations in a real system and used the data to verify the effectiveness of the reconstructed intensity of the proposed method. The implementation system is shown in Fig. 10. Both the grayscale and color cameras (Basler acA1920-155 um and Basler acA1920-155 uc, respectively) with IMX174 grayscale CMOS sensors manufactured by Sony delivered 164 frames per second with a 2.3 MP resolution. Both resolutions were 1920 × 1200 and the pixel size was 5.86 μm × 5.86 μm. The coded aperture consisted of a Hadamard matrix made of lithographed chromium etched on a CaF2 optical glass with a pixel pitch of 12 μm. A pixel on the coded aperture corresponded to an approximately 2×2 pixel on the detector manufactured by Shanghai ZhiBan Electronic Science & Technology Co., Ltd. The objective lens were KOWA LM25HC (25mm), the relay lens were KOWA LM50HC (50mm). The dispersive prism was manufactured by Beijing Yongxing Sensing Information Technology Co., Ltd., producing approximately 100-pixel dispersion from 450 nm to 650 nm when the camera binning selected 2 to 1. According to this principle, we calibrated the prototype and obtained the optical properties of the system. To ensure the consistency of the spectral information collected by the two cameras, we placed a Thorlabs FELH0450 Hard-Coated Longpass Filter and a FESH0650 Hard-Coated Shortpass Filter in front of the system, to make a bandpass filter of 450–650 nm. And we used a 50:50 (R:T) Thorlabs Non-Polarizing Beamsplitter Cube Coating for 400 - 700 nm after the bandpass filter.

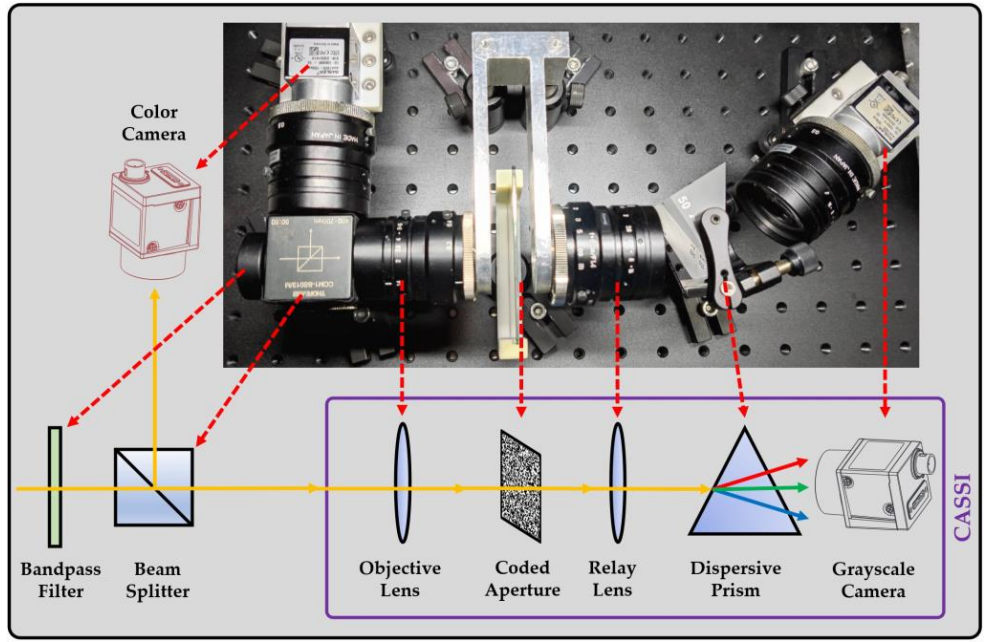

Fig. 10. Schematic illustration of our dual-camera equipment.

We first reconstructed the color boxes, and the results are shown in Fig. 11. The target data consisted of 100 spectral bands with a size of 256 × 355 pixels. The data compressive and color images were captured by the grayscale and color cameras, respectively. Removing images with poor reconstruction quality at both ends can provide at least 80 effective reconstruction images from 450 nm to 650 nm.

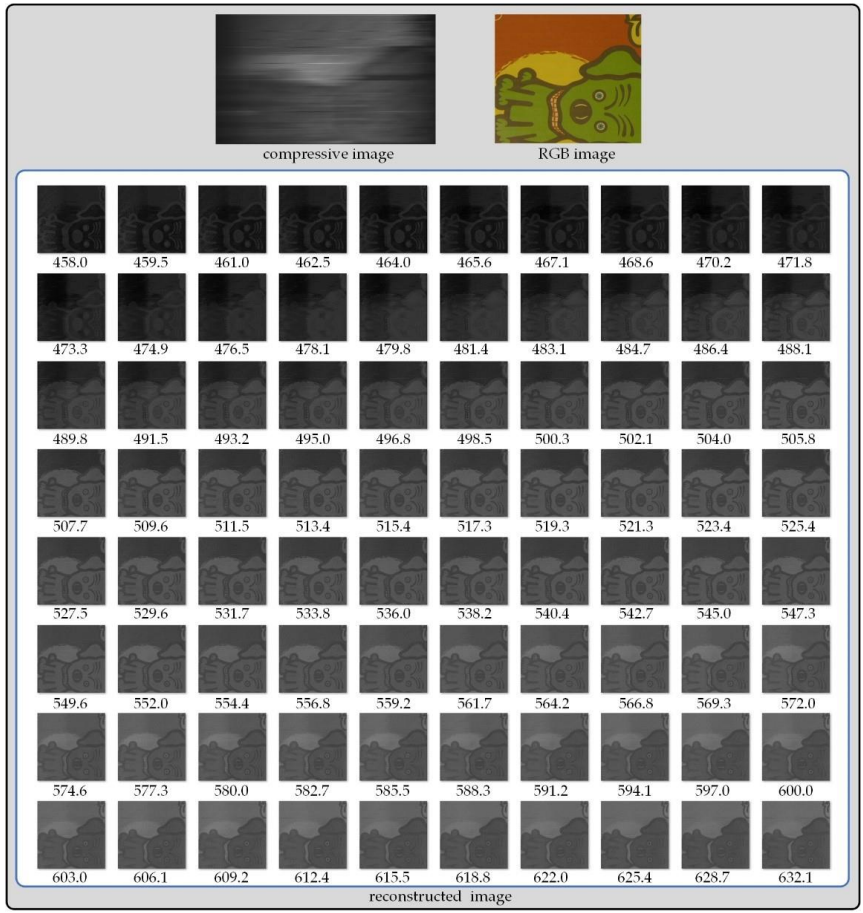

Fig. 11. Reconstructed hyperspectral images of the color box.

We also used the system to image a color card. To obtain the standard data as a reference, we used the visible bandpass filter cut-offs from 450 nm to 650 nm with a step size of 10 nm in front of the grayscale camera to collect the HSIs. Furthermore, we normalized the data with the camera-relative response and bandpass filter transmission data. Fig. 12 shows the five exemplar reconstruction spectra of the standard color card.

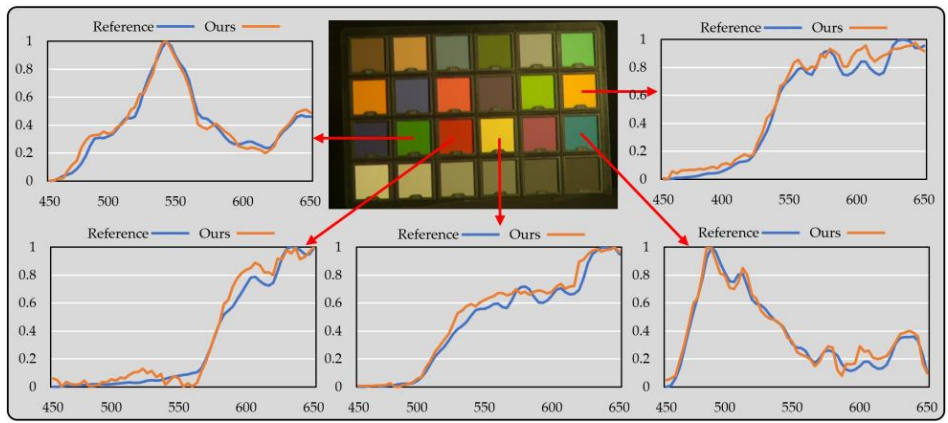

Fig. 12. Five exemplar reconstruction spectra of the standard color card.

## 4. Contributions and Conclusion

In this study, we proposed a self-supervised dual-camera system with untrained neural network for the hyperspectral SCI reconstruction. Which has strong learning capability but doesn't need any training data. Specific contributions are:

(i). Proposed a physics-informed framework with two self-supervised learning branches. One branch projected the physical process of the CASSI to obtain a compressed image to learn the spectral information. Another branch emulated the imaging process of the color camera with the camera's quantum efficiency to learn the spatial information.

(ii). Compared with the iterative optimization method, benefit from the physics-informed CNN framework based on the optical imaging process of the color camera and CASSI, the proposed system exhibited excellent reconstruction performance. Besides, after the optimization, the learned model also has good reconstruction ability for the same type scenarios. Which makes it possible to reconstruct scenes in real time.

(iii). Unlike most supervised deep learning approaches, our method does not require a significant amount of standard data for pre-training. We used the physical imaging process to replace the pathological mapping relationship between the compressive image and the standard HSIs, which resulted in better adaptability to scenes (the situation of no GT).

(iv). Benefiting from physics-informed self-supervised framework, our method can realize online learning, real-time scene generalization and fast and effective imaging in real scenes.

The core of our approach is the introduction of physical constraints. We used the physical process of the CASSI to reconstruct spectral information, and the imaging process of the color camera to replenish spatial information. At the same time, the HSI unmixing idea proposed in this study can be further developed in the field of hyperspectral video acquisition.

**Funding:** This work was supported by the National Natural Science Foundation of China (Grant Nos. 61727802, 61901220, 62101265), Postgraduate Research & Practice Innovation Program of Jiangsu Province (KYCX21_0271), and China Postdoctoral Science Foundation (2021M691592).

**Acknowledgments:** We thank Jiang Yue, Xiaoyu Chen, and Ruobing Ouyang. Jiutao Mu for technical support

**Conflicts of Interest:** The authors declare no conflicts of interest.